%% file: counterfactual_evaluation_regret.tex
\documentclass[sigconf]{acmart}

\usepackage{amsmath,amsfonts}
\usepackage{booktabs}
\usepackage{graphicx}
\usepackage{multirow}
\usepackage{microtype}
\usepackage{tikz}
\usepackage{enumitem}
\usepackage{dsfont} 
\usepackage{algorithm}
\usepackage{algpseudocode}
\usepackage{dsfont}% Math
\usepackage{enumitem}
 
\usepackage{graphicx}
\usepackage{float}
\usepackage{newtxtext,newtxmath}

\usepackage{booktabs}

\usepackage{enumitem}

\usepackage{subcaption}

\usetikzlibrary{arrows.meta,positioning,fit,calc,shapes.geometric}

\renewcommand\footnotetextcopyrightpermission[1]{}

\title{Auditing Marketing Budget Allocation with Hindsight Regret
}

\author{Nilavra Pathak }
\affiliation{
  \institution{ Expedia Group}
  \city{New York}
  \country{USA}
}
\email{npathak@expediagroup.com}

\author{Olivier Jeunen}
\affiliation{
  \institution{aampe}
  \city{Antwerp}
  \country{Belgium}
}
\email{olivier@aampe.com}

\author{ Eric Lambert }
\affiliation{
  \institution{  Expedia Group}
  \city{Seattle}
  \country{USA}
}
\email{elambert@expediagroup.com}
\begin{document}

\begin{abstract}
Organizations routinely make strategic budget allocations under operational constraints. However, they often lack a principled way to assess those realized allocations. We present a retrospective auditing framework based on hindsight regret, for marketing budget allocation. Regret is defined as the opportunity cost of the realized allocation relative to a constraint-faithful benchmark under the same budget and stability guardrails. The framework estimates regime-specific spend--response functions from historical logs, computes feasible hindsight allocations by constrained optimization, and propagates uncertainty through Monte Carlo evaluation to produce regret distributions, expected lift, and probability-of-improvement summaries. This separates allocation inefficiency from uncertainty in the estimated response surfaces. Experiments on real marketing allocation logs show that the framework yields interpretable post-hoc diagnostics and reveals a practical trade-off between allocation flexibility and detectability: moderate feasible re-allocations often capture most measurable gain, while larger shifts move into weak-support regions with higher uncertainty. The result is a practical method for auditing historical budget decisions when online experimentation is costly or infeasible.
\end{abstract}


\maketitle

\input{introduction.tex}
\input{background.tex}
\input{problem_statement}
\input{regret_framework}
\input{results}
\input{Conclusion}
%\newpage
 %\input{regret_adaptive.tex}

\vspace{-0.2cm}

\bibliographystyle{ACM-Reference-Format}
\bibliography{gp_regret_refs}

\newpage

\end{document}

%% file: introduction.tex
\section{Introduction}

Strategic budget allocation is a recurring decision problem in digital advertising. Organizations must distribute finite budget across channels, markets, and campaigns over a planning horizon while respecting operational guardrails such as budget limits, pacing rules, and stability constraints. Because these allocations are high-dimensional, persistent, and costly to revise, poor decisions can carry substantial economic consequences.

A central challenge is retrospective evaluation. Observed outcomes such as revenue or profit reflect not only decision quality, but also market conditions, platform dynamics, and other factors outside the planner's control. As a result, even when historical logs are available, it is difficult to determine whether a realized allocation was close to the best feasible decision that could have been made under the same operational constraints. This raises a natural question: \textit{How much value was left on the table due to the allocation decision itself?}

Online advertising provides a concrete and economically important instance of this problem. Although auctions operate at the level of individual impressions, many consequential choices occur at a slower timescale through portfolio-style budget commitments across heterogeneous assets such as channels, regions, and campaigns \citep{zhao2019unified}. These decisions are subject to practical constraints including caps, pacing rules, and limits on reallocation across periods \cite{pmlr-v235-balseiro24a, nguyenAdKDD23, holthausen1982advertising, symitsi2022keyword, baardman2019learning}. In practice, organizations often have detailed historical allocation logs, but lack principled tools for auditing whether past allocations were near-optimal.

Standard evaluation approaches do not directly address this need. Randomized experiments are costly or infeasible for continuous, high-dimensional allocations and may violate no-interference assumptions due to cross-channel effects \citep{kohavi2020trustworthy,eckles2017interference}. Pre--post and interrupted time-series analyses are sensitive to non-stationarity and shocks \citep{bernal2017its,angrist2009mhe}. Synthetic control methods rely on stable donor units and cross-sectional comparability assumptions that are difficult to justify for persistent, multi-asset allocation trajectories \citep{abadie2021using,dube2014pooled_techreport}. More broadly, while regret is central in online learning, those formulations focus on sequential adaptation under partial feedback rather than post-hoc auditing of a realized strategic allocation.

We therefore study retrospective budget-allocation auditing through \emph{hindsight regret}: the opportunity cost of the realized allocation relative to the best feasible alternative in hindsight under the same operational guardrails. Our goal is not online control or policy learning, but post-hoc measurement of whether a historical allocation left recoverable value unrealized. To do so, we estimate regime-specific spend--response functions from historical logs, construct constraint-faithful hindsight benchmarks through constrained optimization, and propagate model and outcome uncertainty through Monte Carlo evaluation to obtain regret distributions and probability-of-improvement summaries. This yields an interpretable estimate of ``\textit{value left on the table}'' while separating apparent allocation inefficiency from uncertainty in the estimated response surfaces.

\noindent\textbf{In summary, our main contributions are:}
\begin{enumerate}[topsep=0pt,itemsep=0pt,parsep=0pt]
  \item \textbf{A hindsight-regret formulation for retrospective allocation auditing.}
  We cast post-hoc evaluation of strategic budget decisions as comparison against a constraint-faithful hindsight benchmark, yielding an interpretable measure of opportunity cost.

  \item \textbf{A practical offline auditing framework.}
  We combine regime-specific spend--response estimation, constrained hindsight optimization, and uncertainty propagation to evaluate historical allocations using logged data alone.

  \item \textbf{An empirical study on real strategic allocation logs.}
  We show how regret distributions, detectability thresholds, and stability constraints together support systematic post-hoc assessment of budget decisions.
\end{enumerate}

%% file: background.tex
\section{Background \& Related Work}
\label{sec:background}

\textbf{Overview of Measurement Approaches in Practice:}
Randomized online experiments remain the standard for causal
measurement in digital systems \citep{kohavi2020trustworthy}. However,
strategic budget allocation is continuous, high-dimensional, and
persistent, which makes large-scale randomization costly and often
operationally infeasible. Advertising systems also exhibit interference
across auctions, channels, and users, weakening standard no-interference
assumptions \citep{eckles2017interference}.
When experimentation is unavailable, practitioners often fall back on
pre--post or interrupted time-series designs
\citep{bernal2017its,angrist2009mhe}, which rely on stable untreated
dynamics and can break under demand shocks or non-stationarity.
Synthetic control methods similarly require donor stability and
cross-sectional comparability
\citep{abadie2021using,dube2014pooled_techreport,zhang2019inference}, assumptions that are difficult
to justify for persistent, policy-driven allocation trajectories in
mediated ad-delivery systems.

\textbf{Regret as retrospective decision evaluation:}
In statistical decision theory, regret formalizes an \emph{ex-post
opportunity cost}: given realized outcomes and a feasible action set,
the chosen decision is evaluated against the best feasible alternative
that could have been selected in hindsight under the same environment
\citep{savage1954foundations,loomes1982regret}. This produces a static
optimality gap that measures decision inefficiency rather than learning
dynamics. By contrast, regret in bandit and reinforcement learning
settings
\citep{auer2002using,jaksch2010near,srinivas2010gaussian,abbasi2011regret,dean2018regret}
is sequential and cumulative, reflecting the loss incurred while
adaptively learning under partial feedback. Our formulation adopts the
former perspective. It is also distinct from off-policy evaluation,
whose objective is to estimate the value of a target policy from logged
feedback. Here the object of evaluation is a realized allocation
trajectory, and the benchmark is a feasible hindsight reallocation under
the same operational regime rather than a new policy to be deployed
prospectively.

\textbf{Structural Causal Modeling Approach:}
Budget allocation in advertising systems operates through a structural
data-generating process: planned budgets affect latent bidding and
delivery decisions, which in turn determine realized spend and
downstream returns \citep{bottou2013counterfactual}. We use the
Structural Causal Model (SCM) framework \citep{pearl2009causality} to
give semantics to these counterfactual comparisons through
interventional response functions, namely expected returns under
hypothetical spend allocations. Identifiability of such structural
response functions requires assumptions on functional form and noise
structure. Prior work on additive and related noise models provides
identifiability results for nonlinear cause--effect pairs
\citep{shimizu2006linear,hoyer2009nonlinear,peters2017elements,
immer2023identifiability,yin2024effective}.

\textbf{Need for a Custom Framework:}
Existing approaches do not directly answer the question central to this
paper: whether a realized strategic allocation was close to the best
\emph{feasible} allocation that could have been chosen in hindsight
under the same operational guardrails. Our framework addresses this gap
by estimating regime-specific response surfaces from historical logs,
constructing feasible hindsight reallocations within the same regime,
and quantifying the resulting opportunity cost with uncertainty-aware
regret summaries. Non-stationarity is handled through epoch-wise
re-estimation and support-aware uncertainty inflation rather than by
assuming invariant untreated baselines or stationary learning dynamics.

%% file: problem_statement.tex
\section{Problem Formulation}
\label{sec:problem_formulation}

We consider sequential budget allocation over a fixed planning horizon of
$T$ operational time steps. Budgets are updated at coarser decision epochs
$\tau \in \{1,\dots,T/k\}$ and held fixed for $k$ operational steps.
At each epoch, the planner allocates budget across $N$ assets.

Let $\mathbf{b}_\tau=(b_{\tau,1},\dots,b_{\tau,N})
\in \mathcal{B}\subseteq\mathbb{R}_+^N$
denote the epoch-$\tau$ budget vector, where $\mathcal{B}$ encodes feasibility
constraints such as total budget, caps, floors, and operational guardrails.

\paragraph{\textbf{Mediated execution.}}
Budgets do not directly determine spend or returns. Instead, delivery is mediated
by a platform execution mechanism interacting with a stochastic market
environment~\cite{bottou2013counterfactual}. Let $x_\tau$ denote observed context,
$u_\tau$ unobserved exogenous factors (e.g., competition, inventory), and
$z_\tau$ internal system state. The structural relations are
\begin{align}
\boldsymbol{\beta}_\tau
&= \pi(\mathbf{b}_\tau, x_\tau, z_\tau, u_\tau), \\
\mathbf{s}_\tau
&= \sigma(\boldsymbol{\beta}_\tau, x_\tau, z_\tau, u_\tau)
+ \boldsymbol{\xi}_\tau, \quad
\mathbb{E}[\boldsymbol{\xi}_\tau|\cdot]=0, \\
\boldsymbol{\rho}_\tau
&= \rho(\mathbf{s}_\tau, x_\tau, u_\tau)
+ \boldsymbol{\varepsilon}_\tau, \quad
\mathbb{E}[\boldsymbol{\varepsilon}_\tau|\cdot]=0.
\end{align}
Here $\mathbf{s}_\tau$ denotes realized spend and
$\boldsymbol{\rho}_\tau$ realized returns. Latent bidding and serving actions
$\boldsymbol{\beta}_\tau$ are not observed. Our retrospective analysis therefore
operates at the realized-spend layer, treating the upstream execution mechanism
as latent.

\paragraph{\textbf{Observability and feedback.}}
At epoch $\tau$ the planner observes
$(x_\tau,\mathbf{b}_\tau,\mathbf{s}_\tau,\boldsymbol{\rho}_\tau)$,
but not $\boldsymbol{\beta}_\tau$, $z_\tau$, $u_\tau$, or counterfactual outcomes.
Feedback is bandit-style, revealing returns only at the executed allocation.

\paragraph{\textbf{Budget accounting.}}
The planner is given a fixed total budget $B_{\mathrm{tot}}$ for the horizon,
with allocations satisfying
\begin{equation}
\sum_{\tau=1}^{T/k}\sum_{i=1}^N b_{\tau,i}
=
B_{\mathrm{tot}},
\qquad
\mathbf{b}_\tau \in \mathcal{B} \ \forall \tau.
\label{eq:budget_accounting}
\end{equation}

\paragraph{\textbf{Planning policy.}}
A policy $\Pi$ maps available information to allocations:
$
\mathbf{b}_\tau \leftarrow \Pi(\mathcal{F}_\tau)
$,
with
$
\mathcal{F}_\tau =
\{x_{1:\tau}, \mathbf{b}_{1:\tau-1},
\mathbf{s}_{1:\tau-1}, \boldsymbol{\rho}_{1:\tau-1}\}.
$
This partially observed sequential formulation defines the historical setting
against which hindsight regret is evaluated.

%% file: regret_framework.tex
\section{Retrospective Regret Framework}
\label{sec:regret_framework}

We present a retrospective framework for auditing the quality of a
realized budget allocation through \emph{hindsight regret}, defined as
the opportunity cost relative to the best feasible spend trajectory
that could have been executed under the same operational constraints
and historical regime. The pipeline proceeds in three steps:
\begin{enumerate}[topsep=0pt,itemsep=2pt,leftmargin=1.2em]
    \item \textbf{Interventional spend--return modeling (Step I):} 
    Estimate uncertainty aware interventional spend--return response
    functions under a reduced-form grey-box abstraction.

    \item \textbf{Constraint-faithful oracle allocation (Step II):}
    Compute the best feasible benchmark spend trajectory under
    operational guardrails via constrained optimization.

    \item \textbf{Uncertainty-aware regret estimation (Step III):}
    Propagate response uncertainty by Monte Carlo simulation to obtain
    a distribution over regret and associated summary metrics.
\end{enumerate}

\subsection{Empirical regret}
\label{sec:empirical_regret}

In classical decision theory, \emph{regret} is the loss in utility
incurred by a decision that is suboptimal relative to an alternative
evaluated in hindsight under the realized state of the world
\citep{savage1954foundations,loomes1982regret}. In our setting, a
decision corresponds to a feasible spend trajectory
$\mathbf{s}=\{s_{\tau,i}\}_{\tau,i}$ and utility is total realized
return
$R(\mathbf{s}) := \sum_{\tau,i} \rho_{\tau,i}(s_{\tau,i})$.
Given a realized trajectory $\mathbf{s}_{\mathrm{real}}$ and a feasible
benchmark $\mathbf{s}^\star$, regret is defined as
\begin{equation}
\label{eq:regret_def}
\mathrm{Reg}
:=
R(\mathbf{s}^\star) - R(\mathbf{s}_{\mathrm{real}}).
\end{equation}

Because realized returns are stochastic and response functions are
estimated from data, $\mathrm{Reg}$ is a random variable. Individual
realizations may be negative, corresponding to favorable stochastic
outcomes under the realized trajectory relative to the benchmark. We
therefore evaluate \emph{expected regret}
\begin{equation}
\label{eq:expected_regret}
\mathbb{E}[\mathrm{Reg}]
=
\mathbb{E}\!\left[
R(\mathbf{s}^\star) - R(\mathbf{s}_{\mathrm{real}})
\right],
\end{equation}
and report probability statements of the form
$\mathbb{P}(\mathrm{Reg} > 0)$, which quantify the likelihood that the
benchmark dominates the realized trajectory in hindsight.

\paragraph{\textbf{Operational regret with stability constraints ($\delta$) and confidence ($\varepsilon$).}}
The benchmark depends on the operational realism encoded by the
feasible set. We index the feasible set by a stability parameter
$\delta$ and define
\begin{equation}
\label{eq:regret_delta}
\mathrm{Reg}(\delta)
:=
R(\mathbf{s}^\star(\delta)) - R(\mathbf{s}_{\mathrm{real}}),
\qquad
\mathbf{s}^\star(\delta)\in
\arg\max_{\mathbf{s}\in\mathcal{B}(\delta)} R(\mathbf{s}),
\end{equation}
where $\mathcal{B}(\delta)$ denotes the $\delta$-constrained feasible
set (defined in Step~II). We call regret \emph{$\varepsilon$-certified}
at level $\delta$ if
\begin{equation}
\label{eq:epsilon_cert}
\mathbb{P}\!\left(\mathrm{Reg}(\delta) > 0\right) \ge \varepsilon.
\end{equation}
Let $\Delta$ denote a prespecified set of plausible constraint levels.
We select
\begin{equation}
\label{eq:delta_star_def}
\delta^\star
\in
\arg\max_{\delta\in\Delta}
\mathbb{E}\!\left[\mathrm{Reg}(\delta)\right]
\quad \text{s.t.} \quad
\mathbb{P}\!\left(\mathrm{Reg}(\delta) > 0\right) \ge \varepsilon,
\end{equation}
and report $\mathbb{E}[\mathrm{Reg}(\delta^\star)]$ together with
$\mathbb{P}(\mathrm{Reg}(\delta^\star)>0)$.

\begin{table*}[!t]
\centering
\footnotesize
\begin{tabular}{p{2.5cm} p{5.2cm} p{8.8cm}}
\toprule
\textbf{Metric} & \textbf{Definition} & \textbf{Interpretation} \\
\midrule
Expected regret
&
$\widehat{\mathbb{E}}[\mathrm{Reg}(\delta^\star)]
= \frac{1}{J}\sum_{j=1}^{J}\mathrm{Reg}^{(j)}(\delta^\star)$
&
Average achievable gain from feasible hindsight reallocation at the
selected constraint level. \\

Standard deviation
&
$\widehat{\sigma}_{\mathrm{Reg}}
= \sqrt{\frac{1}{J-1}\sum_{j}
(\mathrm{Reg}^{(j)}(\delta^\star) -
\widehat{\mathbb{E}}[\mathrm{Reg}(\delta^\star)])^2}$
&
Sensitivity of regret to outcome noise and model uncertainty; large
values indicate fragile conclusions. \\

Credible interval
&
$\mathrm{CI}_{1-\alpha}
= [Q_{\alpha/2}, Q_{1-\alpha/2}]$
&
Range of plausible regret values under uncertainty; intervals spanning
zero indicate ambiguous dominance. \\

Probability of improvement
&
$\widehat{\mathbb{P}}(\mathrm{Reg}(\delta^\star) > 0)
= \frac{1}{J}\sum_{j}\mathds{1}\{\mathrm{Reg}^{(j)}(\delta^\star) > 0\}$
&
Likelihood that the $\delta^\star$-constrained benchmark outperforms the
realized trajectory; compared against $\varepsilon$ in
\eqref{eq:epsilon_cert}. \\
\bottomrule
\end{tabular}
\captionsetup{font=small}
\caption{Empirical regret metrics and uncertainty summaries. All
quantities are computed from Monte Carlo draws
$\{\mathrm{Reg}^{(j)}(\delta)\}_{j=1}^J$. Unless otherwise stated, we
evaluate at $\delta=\delta^\star$ selected by
\eqref{eq:delta_star_def}.}
\label{tab:regret_metrics}
\end{table*}

\subsection{Step I: Grey-Box Response Functions}
\label{sec:gb_identification}

We model epoch-specific spend--return mappings and evaluate
counterfactual spend trajectories under fixed operational guardrails.
Upstream budget-to-bid dynamics are abstracted away; analysis is
conducted at the realized-spend layer.

\subsubsection{\textbf{Structural Target and Identification}}
\label{sec:gb_assumptions}

We formalize the assumptions used to identify an epoch-specific
interventional mean response. The key modeling choice is to treat each
epoch as a locally stationary regime: within an epoch, observations are
analyzed as conditionally independent replicates given observed
covariates, while non-stationarity is handled at epoch boundaries
through drift regularization and stability constraints.

\paragraph{\textbf{A1. Structural Response Model.}}
For asset $i$ in epoch $e$, realized outcomes satisfy
\begin{equation}
\label{eq:structural_model_epoch}
\rho_{\tau,i}
=
f_{i,e}(s_{\tau,i}, x_{\tau,i})
+
\varepsilon_{\tau,i},
\qquad \tau\in e.
\end{equation}
The function $f_{i,e}$ represents the reduced-form spend--return
mechanism induced by the execution regime in epoch $e$.

\paragraph{\textbf{A2. Within-epoch stationarity and local mean independence (identification).}}
Within epoch $e$, conditional on $(s_{\tau,i},x_{\tau,i})$, the noise
process $\{\varepsilon_{\tau,i}\}_{\tau\in e}$ is treated as mean-zero
and approximately independent across $\tau$:
\begin{equation}
\label{eq:mean_independence_epoch}
\mathbb{E}[\varepsilon_{\tau,i}\mid s_{\tau,i}, x_{\tau,i}]
=
0,
\qquad
\varepsilon_{\tau,i}\ \perp\!\!\!\perp\ \varepsilon_{\tau',i}
\ \ (\tau\neq\tau';\ \tau,\tau'\in e).
\end{equation}
Under A2, the interventional mean response
\[
\mu_{i,e}(s)
=
\mathbb{E}\!\left[f_{i,e}(s, X_{\tau,i}) \mid \tau\in e\right]
\]
is identified over the \emph{realized spend range} in epoch $e$, that
is, the $s$ values observed under the historical policy within that
epoch.

\paragraph{\textbf{A3. Bounded drift across epochs (regularization only).}}
Across adjacent epochs $e$ and $e'$, the reduced-form mechanism varies
smoothly on the overlap of their realized spend ranges:
\begin{equation}
\label{eq:bounded_drift}
\sup_{s\in\mathcal S_{e,e'}}
|f_{i,e}(s)-f_{i,e'}(s)| \le \delta,
\qquad
\mathcal S_{e,e'} := \mathrm{range}(e)\cap \mathrm{range}(e').
\end{equation}
This condition enables weak borrowing of strength for estimation
stability but does not expand causal identification beyond the epoch's
realized spend range. Accordingly, the framework is intended for
within-regime retrospective auditing, with conservative
uncertainty-aware evaluation when optimization reaches weak-support
regions.

\subsubsection{\textbf{Drift-regularized grey-box model}}
\label{sec:gb_model}

The mean response is decomposed as
\begin{equation}
\label{eq:greybox_model}
\rho_{\tau,i}
=
\mu_{\mathrm{nom},i,e}(s_{\tau,i};\theta_{i,e})
+
g_{i,e}(s_{\tau,i})
+
\sigma_{i,e}(s_{\tau,i})\,\varepsilon_{\tau,i}.
\end{equation}

\paragraph{\textbf{Nominal saturation.}}
\begin{equation}
\label{eq:nominal_saturation}
\mu_{\mathrm{nom},i,e}(s)
=
a_{i,e}(1-e^{-b_{i,e}s}),
\qquad a_{i,e},b_{i,e}\ge0.
\end{equation}

\paragraph{\textbf{Residual mean and variance.}}
\[
g_{i,e}(\cdot)\sim\mathcal{GP}(0,k_g),
\qquad
\log\sigma^2_{i,e}(\cdot)\sim\mathcal{GP}(m_\sigma,k_\sigma),
\]
with Mat\'ern$(2.5)$ kernels.

The two Gaussian processes serve distinct purposes. The mean GP
$g_{i,e}$ captures systematic deviations from the nominal saturation
curve, allowing flexible local curvature while preserving the
interpretable global shape imposed by
$\mu_{\mathrm{nom},i,e}$. The variance GP models heteroskedastic
outcome dispersion as a function of spend, enabling uncertainty to
increase in weak-support regions. Separating mean and variance
components yields calibrated counterfactual uncertainty, which is
essential for Monte Carlo regret estimation and detectability analysis.

\paragraph{\textbf{Auxiliary regularization.}}
Neighboring-epoch samples are incorporated with exponential weights
\[
w_t=\alpha_{\mathrm{aux}}\exp(-|d_t|/\tau),
\]
which is equivalent to treating auxiliary points as noisier
observations. Core epoch data determine identification; auxiliary data
provide weak curvature regularization under bounded drift.

\begin{figure}[t]
    \centering
    \begin{subfigure}[t]{0.48\linewidth}
        \centering
        \includegraphics[width=\linewidth]{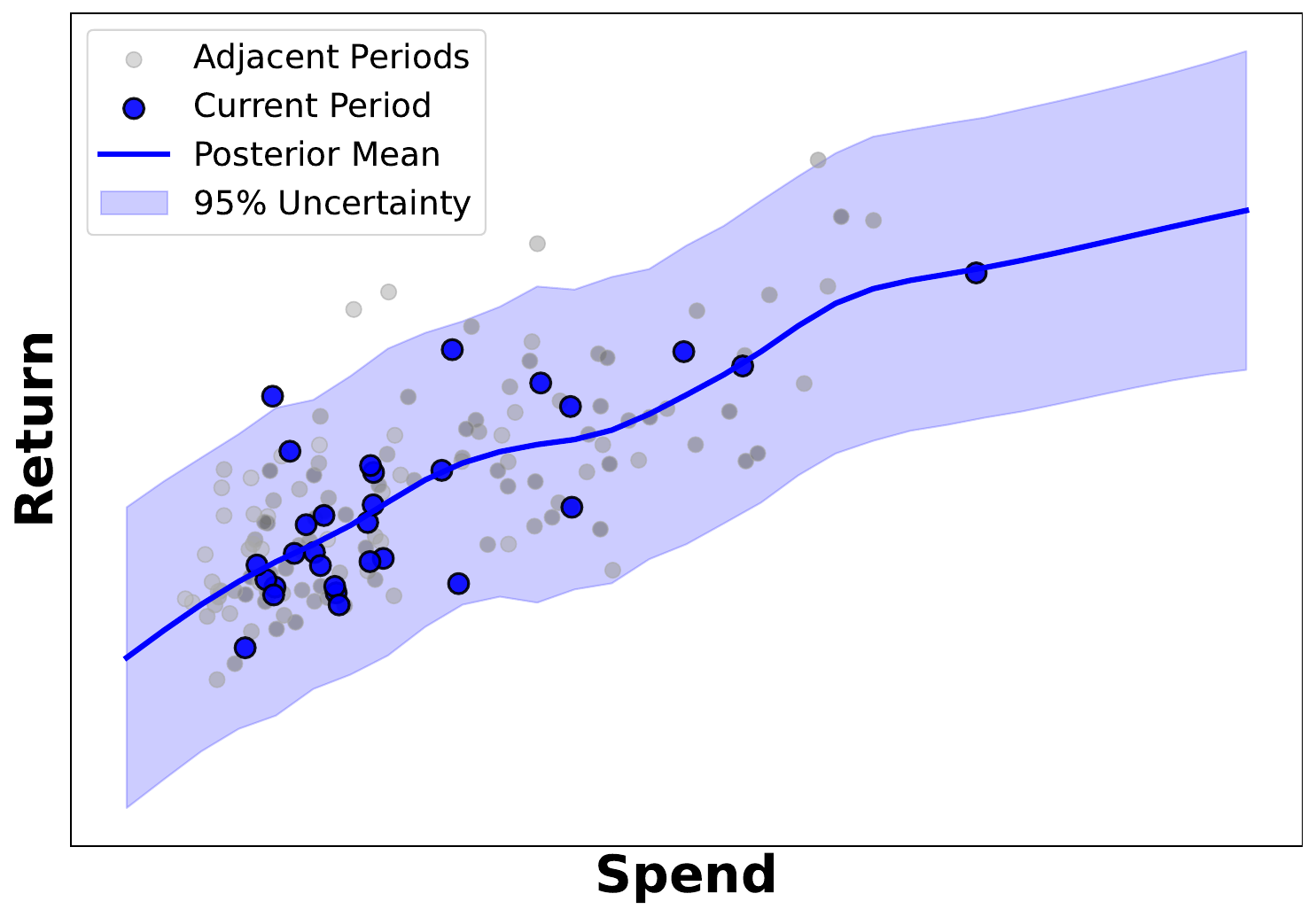}
    \end{subfigure}
    \hfill
    \begin{subfigure}[t]{0.48\linewidth}
        \centering
        \includegraphics[width=\linewidth]{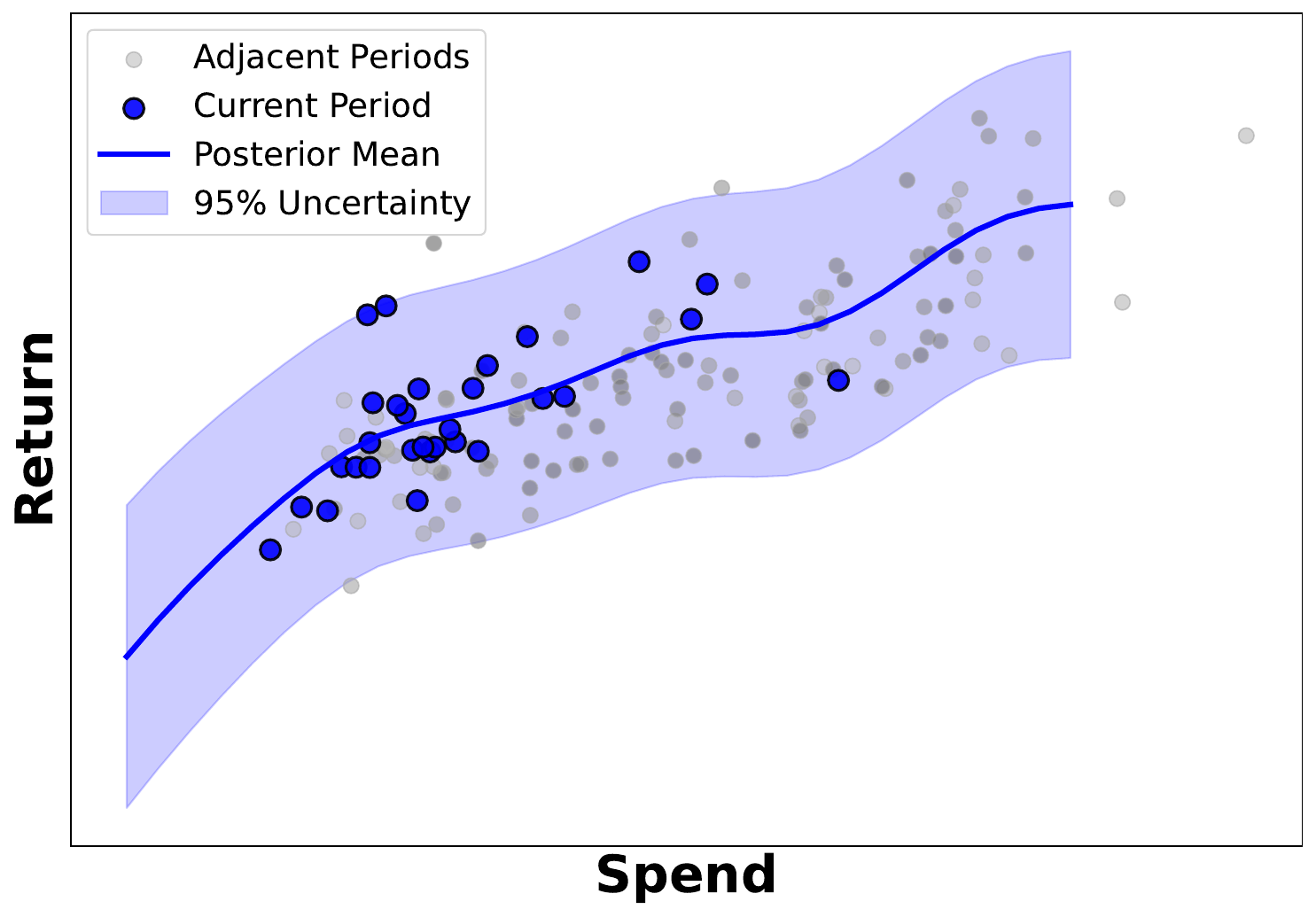}
    \end{subfigure}
    \captionsetup{font=small}
    \caption{Monotone grey-box spend--response estimation under bounded
    temporal drift. Current-period observations define the core realized
    spend range (blue), while adjacent periods (grey) provide weakly
    weighted curvature information. Posterior uncertainty widens
    outside the core range to reflect increased epistemic risk.}
    \label{fig:greybox_examples}
\end{figure}

\paragraph{\textbf{Support-aware extrapolation.}}
For $s$ beyond the upper boundary of the core epoch's realized spend
range, the fitted mean is projected onto the space of non-decreasing
functions via boundary-anchored isotonic regression. For $s$ beyond the
weighted realized range (core plus down-weighted auxiliary coverage),
epistemic variance is inflated:
\begin{equation}
\label{eq:epi_inflation}
\sigma_{\mathrm{epi}}^2(s)
=
\sigma_g^2(s)
+
\left[
\kappa_{\mathrm{right}}
\left(
\frac{\max\{0,s-s_{\mathrm{hi}}^{\mathrm{all}}\}}
{s_{\mathrm{scale}}}
\right)^p
\sigma_{\mathrm{res}}
\right]^2.
\end{equation}
These regions are used only for conservative regret evaluation. Within
the core realized range, monotonicity is not imposed; in practice, the
posterior mean is typically monotone due to the dominance of the
nominal saturation component and smooth GP regularization. Example of the constructed oracle functional forms are shown in Figure~\ref{fig:greybox_examples}.

\subsection{Step II: Oracle allocation under operational feasibility constraints}
\label{sec:oracle_opt}

Given the fitted epoch-wise response functions
$\{\hat{\mu}_{i,e}(\cdot)\}_{i=1}^{K}$, we define an oracle allocation
as the solution to a fixed objective evaluated over alternative feasible
sets:
\begin{equation}
\label{eq:oracle_objective}
\max_{\{s_{\tau,i}\}_{\tau,i}}
\quad
\sum_{e=1}^{E}\sum_{i=1}^{K}\hat{\mu}_{i,e}(s_{e,i}),
\end{equation}
where $s_{e,i}$ denotes the spend assigned to asset $i$ in epoch $e$.
The distinction between oracle variants arises solely from differences
in the \emph{feasible set}, which encodes alternative assumptions about
operational realism.

\subsubsection{\textbf{Unconstrained oracle (idealized hindsight)}}
\label{sec:oracle_unconstrained}

The unconstrained oracle evaluates \eqref{eq:oracle_objective} over the
largest feasible set $\mathcal{B}_{\mathrm{unc}}$, defined only by
budget accounting and asset-level bounds:
\[
\mathcal{B}_{\mathrm{unc}}
:=
\left\{
\mathbf{s}
\;\middle|\;
\sum_{e,i} s_{e,i} = B_{\mathrm{tot}},\;
\underline{s}_i \le s_{e,i} \le \overline{s}_i
\right\}.
\]

This oracle represents a theoretical upper bound corresponding to
perfect hindsight and unrestricted reallocation across epochs.

\subsubsection{\textbf{Constrained oracle (realism-adjusted)}}
\label{sec:oracle_constrained}

Unconstrained optimization may produce trajectories that are
operationally implausible. To restore realism, we introduce a
constrained family
$\mathcal{B}_{\mathrm{con}}(\delta)\subset \mathcal{B}_{\mathrm{unc}}$
that limits how allocations may change across decision epochs. We
impose epoch-to-epoch stability constraints:
\begin{equation}
\label{eq:mom_constraint_epoch}
s_{e-1,i}(1-\delta)\le s_{e,i}\le s_{e-1,i}(1+\delta),
\qquad \forall e>1,\ i,
\end{equation}
where $\delta\in(0,1)$ encodes allowable relative changes.

The constrained oracle is defined as
\[
\mathbf{s}^{\star}_{\mathrm{con}}(\delta)
\in
\arg\max_{\mathbf{s}\in\mathcal{B}_{\mathrm{con}}(\delta)}
\sum_{e=1}^{E}\sum_{i=1}^{K}\hat{\mu}_{i,e}(s_{e,i}).
\]

\begin{figure*}[t]
    \centering
    \footnotesize
    \begin{subfigure}[t]{0.32\textwidth}
        \centering
        \includegraphics[width=\linewidth]{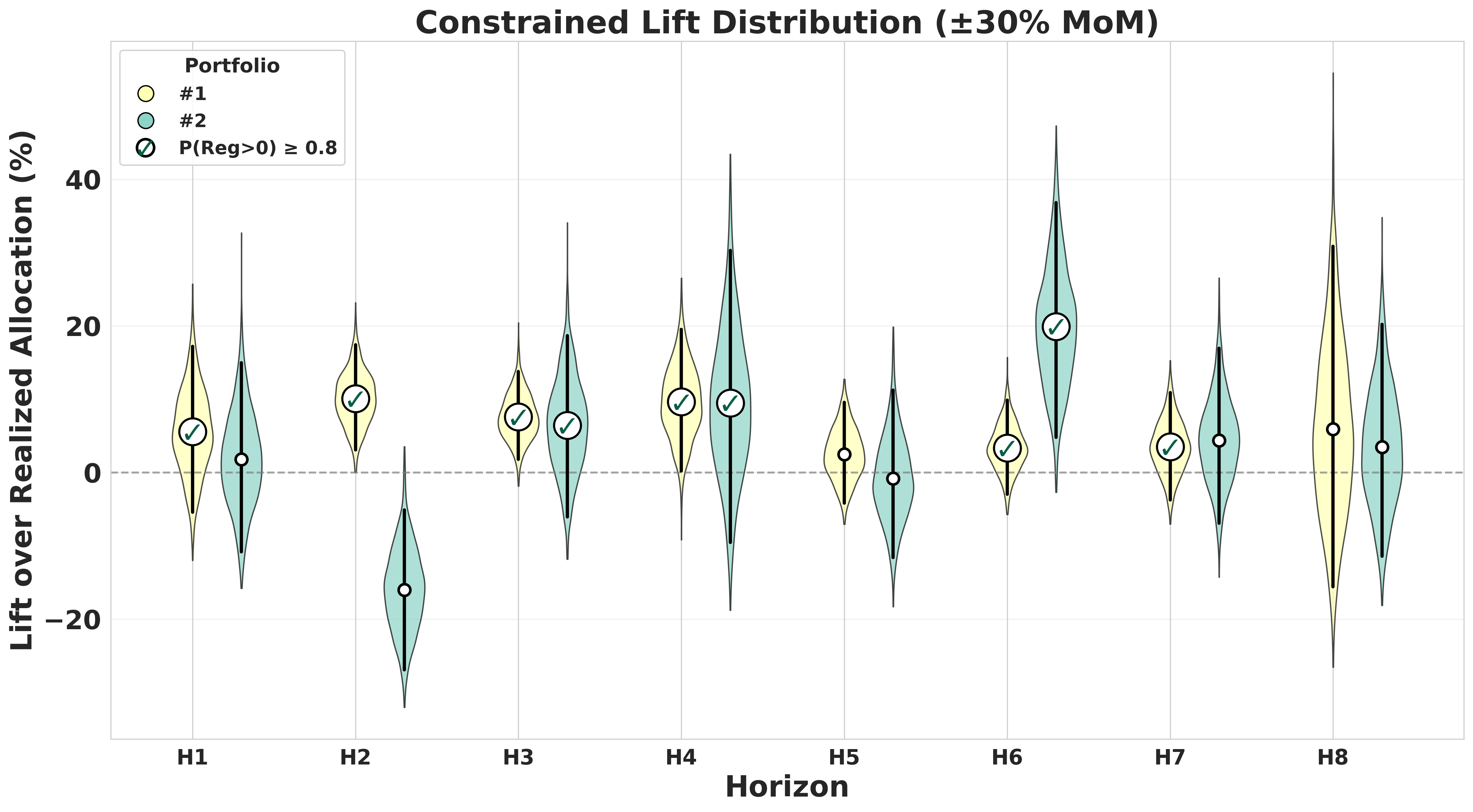}
        \captionsetup{font=footnotesize}
        \caption{$\delta = 30\%$}
        \label{fig:lift_rho30}
    \end{subfigure}
    \hfill
    \begin{subfigure}[t]{0.32\textwidth}
        \centering
        \includegraphics[width=\linewidth]{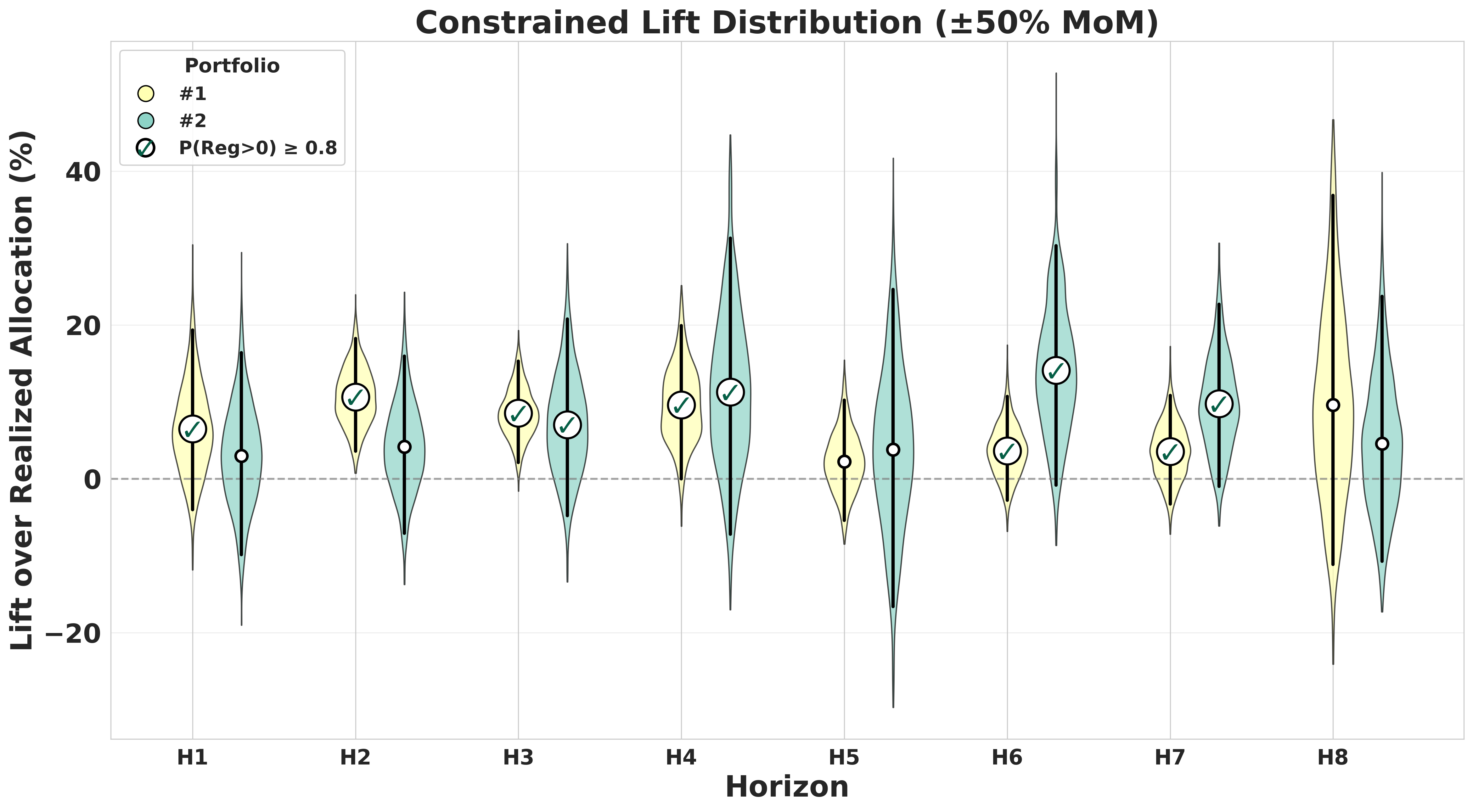}
        \captionsetup{font=footnotesize}
        \caption{$\delta = 50\%$}
        \label{fig:lift_rho50}
    \end{subfigure}
    \hfill
    \begin{subfigure}[t]{0.32\textwidth}
        \centering
        \includegraphics[width=\linewidth]{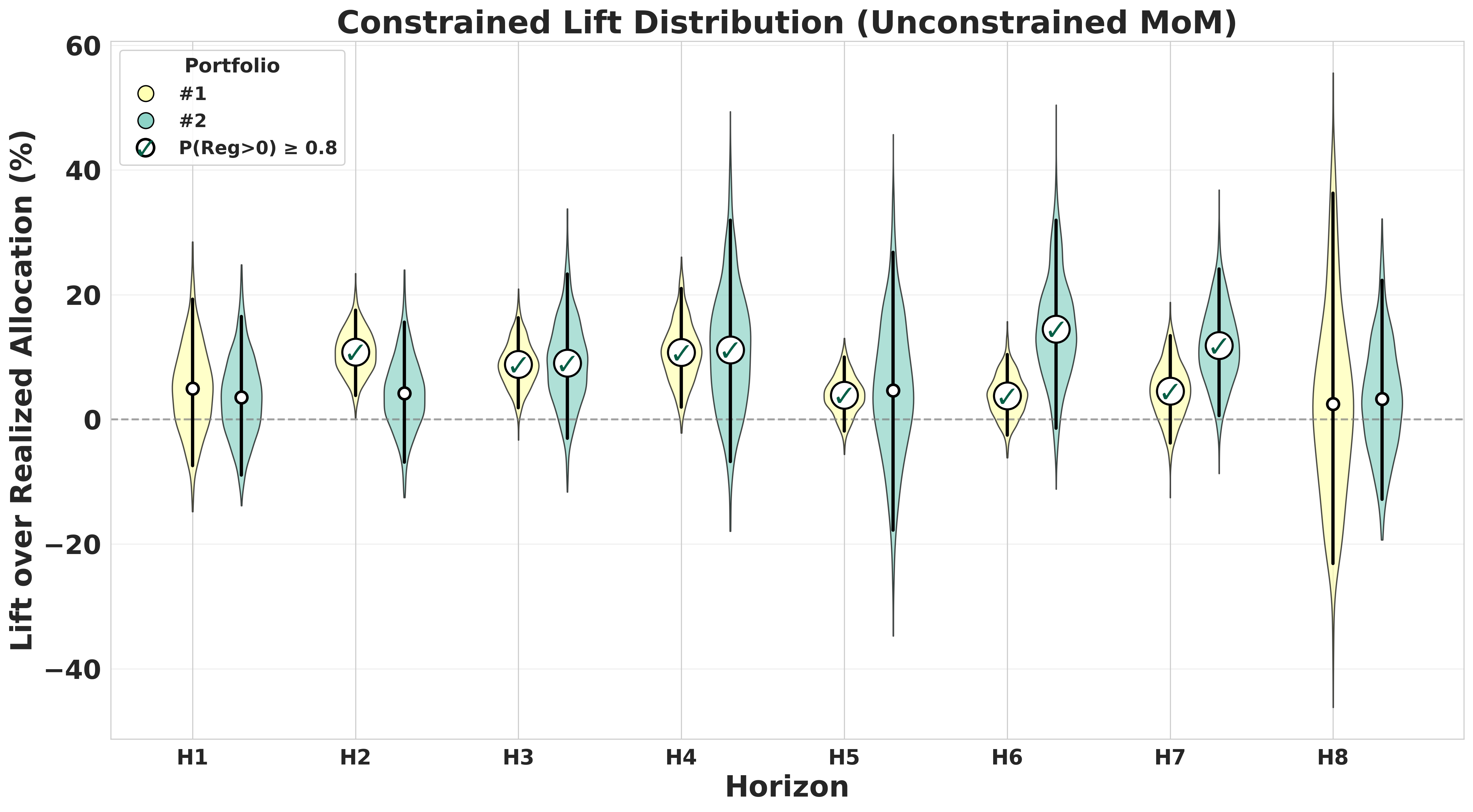}
        \captionsetup{font=footnotesize}
        \caption{Unconstrained}
        \label{fig:lift_uncon}
    \end{subfigure}
    \captionsetup{font=footnotesize}
    \caption{Monte Carlo regret (lift) distributions under alternative
    stability constraints. Each violin summarizes posterior lift
    relative to the realized allocation across portfolio--horizon pairs.
    Increasing flexibility expands the feasible set but does not
    guarantee monotone improvements in uncertainty-aware lift.}
    \label{fig:constrained_lift_comparison}
\end{figure*}

\subsubsection{\textbf{Relationship between oracle variants under support-aware planning}}
\label{sec:oracle_relationship}

Oracle variants differ only through the feasible set
$\mathcal{B}(\delta)$, where larger $\delta$ permits greater
epoch-to-epoch reallocation. For the deterministic planning objective
$\tilde R(\mathbf{s})$ in Step~II, nested feasibility implies
\[
\max_{\mathbf{s}\in\mathcal{B}(\delta_2)} \tilde R(\mathbf{s})
\;\ge\;
\max_{\mathbf{s}\in\mathcal{B}(\delta_1)} \tilde R(\mathbf{s}),
\quad \delta_2>\delta_1.
\]
Thus monotonicity holds for the deterministic planning objective, but
not necessarily for uncertainty-aware empirical regret after posterior
evaluation in Step~III.

However, we report \emph{uncertainty-aware} regret from Step~III, where
the chosen allocation $\mathbf{s}^\star(\delta)$ is evaluated under
posterior predictive uncertainty. Larger $\delta$ may shift the
optimizer toward weakly supported regions, increasing variance and
occasionally reducing both
$\mathbb{E}[\mathrm{Reg}(\delta)]$ and the detectability
$\mathbb{P}(\mathrm{Reg}(\delta)>0)$. Thus operational lift need not be
monotone in $\delta$, and empirical curves may exhibit saturation or
dips.

\subsection{Step III: Regret distribution via Monte Carlo}
\label{sec:mc_regret}

Regret is random due to stochastic outcomes and uncertainty in the
Step~I response surfaces. For each stability level $\delta\in\Delta$
(and the unconstrained case), we first compute the oracle trajectory
$\mathbf{s}^\star(\delta)$ by optimizing the plug-in mean responses from
Step~I over the feasible set from Step~II. We then hold
$\mathbf{s}^\star(\delta)$ fixed and propagate uncertainty via Monte
Carlo to obtain a distribution over $\mathrm{Reg}(\delta)$. For each
scenario we perform the following sampling steps:

\begin{enumerate}[topsep=0pt,itemsep=2pt,leftmargin=1.2em]
    \item \textbf{Sample counterfactual returns.}
    Draw independent innovations
    $\varepsilon_{e,i}^{(j)}\sim\mathcal{N}(0,1)$ and generate
    counterfactual returns using the fitted location--scale model,
    \[
        \rho_{e,i}^{(j)}(s)
        =
        \hat\mu_{i,e}(s)
        + \eta_{i,e}^{(j)}(s)
        + \hat\sigma_{i,e}(s)\,\varepsilon_{e,i}^{(j)},
    \]
    where $\eta_{i,e}^{(j)}(s)$ is a GP-consistent mean perturbation for
    $g_{i,e}$, augmented with the off-range inflation term in
    \eqref{eq:epi_inflation}. This simulation treats draws as
    conditionally independent across assets and epochs given the fitted
    responses. We treat this as a first-order approximation for
    retrospective auditing rather than a full joint model of cross-asset
    dependence.

    \item \textbf{Compute regret at each constraint level.}
    Using the fixed oracle trajectory $\mathbf{s}^\star(\delta)$,
    compute
    \[
        \mathrm{Reg}^{(j)}(\delta)
        =
        R^{(j)}\!\left(\mathbf{s}^\star(\delta)\right)
        -
        R^{(j)}\!\left(\mathbf{s}_{\mathrm{real}}\right),
        \qquad \forall \delta\in\Delta,
    \]
    and analogously
    $\mathrm{Reg}^{(j)}_{\mathrm{unc}}$ for the unconstrained oracle.
\end{enumerate}

We then select $\delta^\star$ via \eqref{eq:delta_star_def} using these
Monte Carlo estimates, including the $\varepsilon$-certificate, and
report the regret distribution at $\delta^\star$ together with the
summary metrics in Table~\ref{tab:regret_metrics}.

%% file: results.tex
\section{Results}
\label{sec:results}

We evaluate the proposed retrospective regret framework using historical
strategic allocation logs under operationally realistic feasibility constraints.
All outcomes are reported as \emph{relative lift} with respect to the realized
allocation. For confidentiality, we omit portfolio composition, asset
identities, absolute magnitudes, and the semantic meaning of evaluation
horizons. Our goal is not to recover a ground-truth global optimum, but to
assess whether the framework yields stable and decision-relevant retrospective
diagnostics under realistic feasibility constraints. We address four questions:
\begin{itemize}[leftmargin=1.5em,topsep=2pt,itemsep=1pt]
    \item \textbf{RQ1:} Do logged allocations support stable and interpretable regret diagnostics?
    \item \textbf{RQ2:} How does uncertainty shape regret?
    \item \textbf{RQ3:} What is the effect of operational stability constraints?
    \item \textbf{RQ4:} When are larger reallocations justified?
\end{itemize}

\begin{figure*}[t]
    \centering 
    \footnotesize
    \begin{subfigure}[t]{0.32\linewidth}
        \centering
        \includegraphics[width=\linewidth]{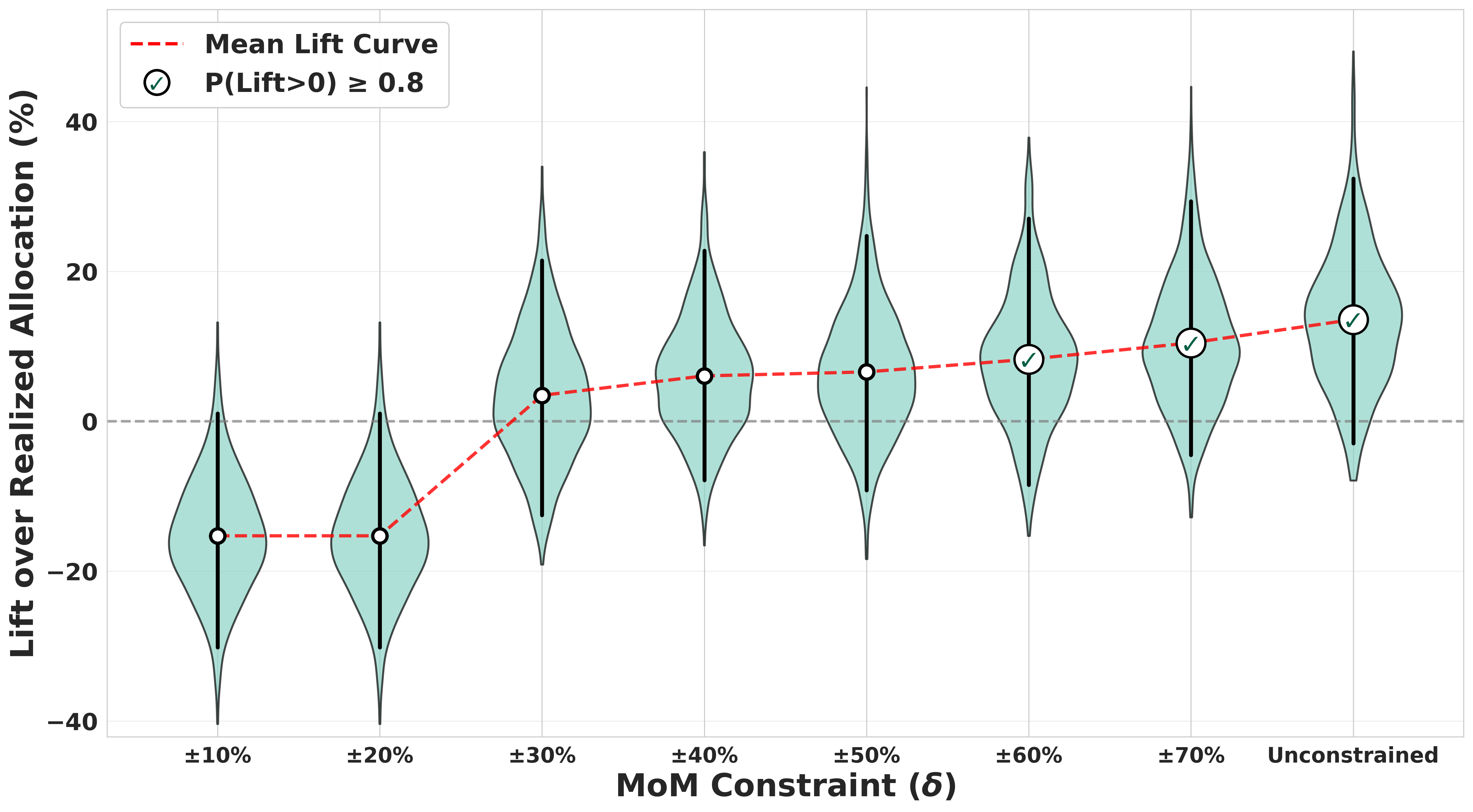} 
        \label{fig:rq4_ablation_a}
    \end{subfigure}
    \hfill
    \begin{subfigure}[t]{0.32\linewidth}
        \centering
        \includegraphics[width=\linewidth]{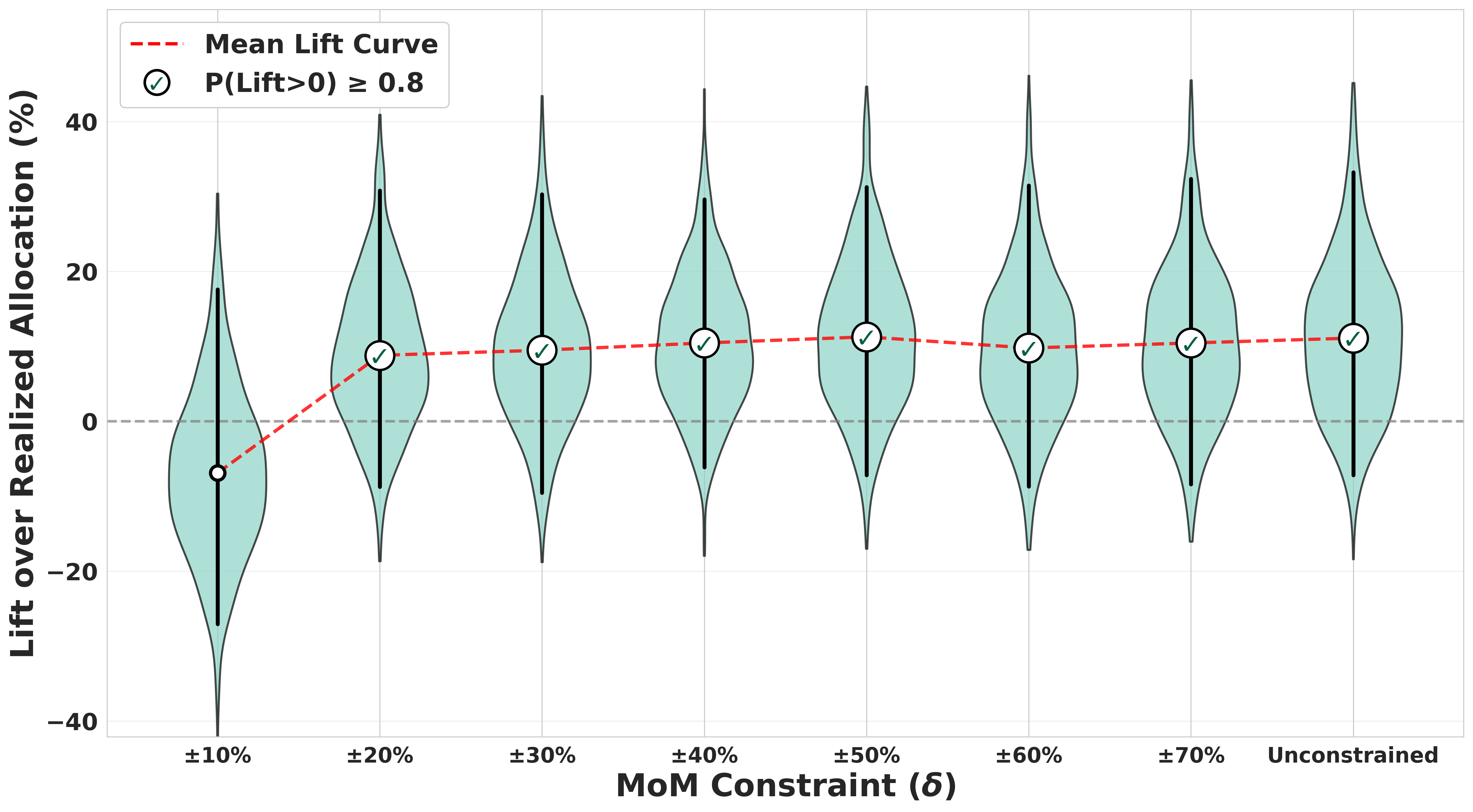} 
        \label{fig:rq4_ablation_b}
    \end{subfigure}
    \hfill
    \begin{subfigure}[t]{0.32\linewidth}
        \centering
        \includegraphics[width=\linewidth]{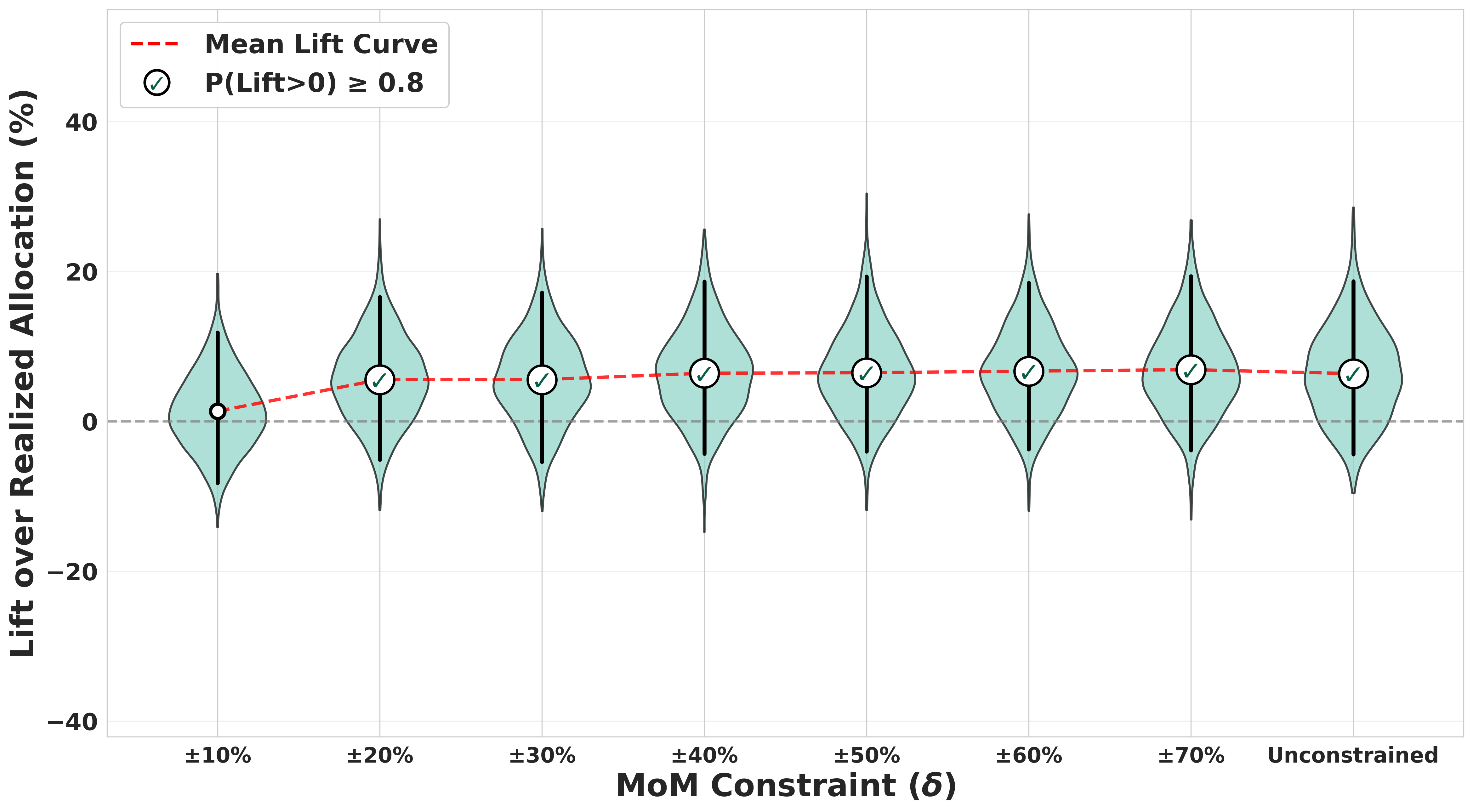} 
        \label{fig:rq4_ablation_c}
    \end{subfigure}
    \captionsetup{font=small}
    \caption{Ablation of stability for three portfolios and horizons. Mean lift
    (dashed line) and lift distributions are shown as a function of allowable
    inter-epoch change \(\delta\). Checkmarks denote
    \(P(\mathrm{Lift}>0)\ge 0.8\). Gains typically concentrate at moderate
    flexibility and saturate thereafter.}
    \label{fig:rq4_ablation_multi}
\end{figure*}

\subsection{Retrospective Regret, Uncertainty, and Feasibility}
\label{sec:rq123}

We evaluate \(P=8\) portfolios across \(H=8\) decision horizons.
For each portfolio--horizon pair, epoch-specific spend--return functions are
estimated using the weighted monotone grey-box model
(Section~\ref{sec:gb_identification}). Oracle allocations are computed once per
feasibility specification and then held fixed. Uncertainty is introduced only at
evaluation time via Monte Carlo propagation of outcome noise and model
uncertainty. Figure~\ref{fig:constrained_lift_comparison} shows lift
distributions for representative portfolios across horizons under multiple
feasibility settings. The distributions illustrate that regret is inherently
distributional and that its dispersion varies across constraint levels.

\textbf{RQ1--RQ3 findings.}
Across all portfolio--horizon pairs we observe:

\begin{enumerate}[leftmargin=1.5em,topsep=2pt,itemsep=2pt]
    \item \textbf{Logged allocations support interpretable regret diagnostics (RQ1).}
    The framework produces well-defined lift distributions and certification
    rates from historical logs without requiring new experimentation, provided
    evaluation remains within or near empirically supported regions.

    \item \textbf{Uncertainty is first-order (RQ2).}
    Many cases exhibit wide lift dispersion even when posterior mean lift is
    positive. Negative realizations arise naturally under Monte Carlo evaluation
    due to stochastic outcomes and epistemic uncertainty.

    \item \textbf{Operational constraints affect both magnitude and detectability (RQ3).}
    Relaxing the stability constraint \(\delta\) typically increases plug-in
    lift, but can widen regret distributions as optimization moves spend toward
    regions with weaker empirical support, reducing high-confidence
    detectability.
\end{enumerate}

\textbf{Aggregate detectability.}
Table~\ref{tab:detectability_rq10_weighted_stable} summarizes certification
results across feasibility levels. For each \(\delta\) and confidence threshold
\(\epsilon\), we report the fraction of portfolio--horizon pairs satisfying
\(P(\mathrm{Reg}>0)\ge\epsilon\) and the mean lift among detectable pairs.

Three consistent patterns emerge. First, moderate thresholds
(\(\epsilon\in\{0.5,0.6\}\)) frequently certify improvement. Second,
stringent thresholds (\(\epsilon\ge0.8\)) substantially reduce detectability,
reflecting broad regret distributions even when mean lift is positive. Third,
increasing flexibility generally raises mean lift among detectable cases but does
not guarantee monotone gains in detectability, consistent with uncertainty
inflation beyond core empirical support. Taken together, these results suggest
that the framework is most useful as an auditing tool for detecting recoverable
opportunity cost under plausible reallocation limits, rather than for
certifying large aggressive reallocations.

\begin{table}[H]
\centering
\scriptsize
\setlength{\tabcolsep}{5pt}
\renewcommand{\arraystretch}{1.15}

\begin{tabular}{l|cccc}
\toprule
\textbf{Constraint $\delta$}
& $\epsilon{=}0.6$
& $\epsilon{=}0.7$
& $\epsilon{=}0.8$
& $\epsilon{=}0.9$ \\
\midrule

$\pm 20\%$
& 0.59 (8.2 $\pm$ 5.6)
& 0.33 (8.3 $\pm$ 5.0)
& 0.20 (8.2 $\pm$ 2.9)
& 0.08 (9.6 $\pm$ 1.5) \\

$\pm 30\%$
& 0.75 (9.4 $\pm$ 6.7)
& 0.53 (10.6 $\pm$ 6.6)
& 0.28 (10.3 $\pm$ 5.3)
& 0.09 (11.4 $\pm$ 4.1) \\

$\pm 40\%$
& 0.88 (9.9 $\pm$ 7.6)
& 0.61 (11.4 $\pm$ 7.6)
& 0.38 (11.3 $\pm$ 7.0)
& 0.09 (12.2 $\pm$ 4.4) \\

$\pm 50\%$
& 0.89 (10.5 $\pm$ 8.0)
& 0.66 (11.9 $\pm$ 8.2)
& 0.42 (12.8 $\pm$ 8.4)
& 0.12 (12.7 $\pm$ 4.9) \\

\midrule
Unconstrained
& 0.89 (12.5 $\pm$ 9.7)
& 0.67 (14.4 $\pm$ 10.0)
& 0.47 (14.0 $\pm$ 9.5)
& 0.22 (14.7 $\pm$ 6.8) \\

\bottomrule
\end{tabular}

\captionsetup{font=footnotesize}
\caption{Fraction of portfolio--horizon pairs with $P(\mathrm{Reg}>0)\ge \epsilon$.
Parentheses show mean $\pm$ std lift (\%) across detectable pairs.}
\label{tab:detectability_rq10_weighted_stable}
\end{table}

 \vspace{-1cm}
\subsection{When Are Large Re-Allocations Justified?}
\label{sec:rq4}

RQ4 evaluates the marginal benefit of relaxing the inter-epoch stability
constraint $\delta$. For each $\delta$, we recompute the oracle allocation and
estimate regret under the same Monte Carlo protocol, isolating the effect of
flexibility alone.

Figure~\ref{fig:rq4_ablation_multi} shows two regimes. Allowing moderate
flexibility ($\pm20\%$ -- $\pm40\%$) yields large gains by correcting obvious
misallocations and shifting spend toward well-supported, high-return regions.
Beyond this, lift often saturates: once allocations approach local optima,
additional movement mainly explores flatter or weakly supported regions where
marginal returns are small and uncertainty is higher.

In some portfolios, lift continues to rise with $\delta$ when profitable regions
remain farther from the realized allocation, so larger moves are still
beneficial. However, greater flexibility typically increases variance, so
detectability improves only selectively and can weaken at stringent confidence
thresholds, as reflected in
Table~\ref{tab:detectability_rq10_weighted_stable}. Overall, most gains are
captured at moderate $\delta$, while extreme reallocations trade reliability for
limited additional return.

%% file: Conclusion.tex
\section{Conclusion and Future Work}

We introduced a hindsight-regret framework for retrospective auditing of
strategic budget allocation under operational constraints. The framework
evaluates realized allocations against constraint-feasible hindsight
benchmarks built from regime-specific response surfaces estimated from
historical logs. Empirically, most recoverable gains are attained through
moderate feasible reallocations, while more aggressive flexibility yields
diminishing returns and higher uncertainty.

Future work will develop formal guarantees for response estimation, regret
perturbation, and finite-sample Monte Carlo summaries. An important next
step is a semi-synthetic benchmark with known response structure and oracle
regret, which would enable direct recovery-based validation of the framework.
We also plan to extend the same setup to policy-level retrospective
scoreboarding of alternative allocation strategies within the same logged
regime.

%% file: gp_regret_refs.bib
@article{holthausen1982advertising,
  title   = {Advertising Budget Allocation under Uncertainty},
  author  = {Holthausen, Duncan M. and Assmus, Gert},
  journal = {Management Science},
  volume  = {28},
  number  = {5},
  pages   = {487--499},
  year    = {1982}
}

@article{symitsi2022keyword,
  title   = {Keyword Portfolio Optimization in Paid Search Advertising},
  author  = {Symitsi, Efthymia and Markellos, Raphael N. and Mantrala, Murali K.},
  journal = {European Journal of Operational Research},
  volume  = {303},
  number  = {2},
  pages   = {767--778},
  year    = {2022}
}

@article{baardman2019learning,
  title={Learning optimal online advertising portfolios with periodic budgets},
  author={Baardman, Lennart and Fata, Elaheh and Pani, Abhishek and Perakis, Georgia},
  journal={Available at SSRN 3346642},
  year={2019}
}

@inproceedings{pmlr-v235-balseiro24a,
  title     = {A Field Guide for Pacing Budget and {ROS} Constraints},
  author    = {Balseiro, Santiago R. and Bhawalkar, Kshipra and Feng, Zhe and Lu, Haihao and Mirrokni, Vahab and Sivan, Balasubramanian and Wang, Di},
  booktitle = {Proceedings of the 41st International Conference on Machine Learning},
  series    = {Proceedings of Machine Learning Research},
  volume    = {235},
  pages     = {2607--2638},
  year      = {2024}
}

@inproceedings{nguyenAdKDD23,
  title     = {Practical Budget Pacing Algorithms And Simulation Test Bed For {eBay} Marketplace Sponsored Search},
  author    = {Nguyen, Ha and Gligorijevic, Djordje and Borah, Arnab and Adalinge, Gajanan and Bagherjeiran, Abraham},
  booktitle = {AdKDD Workshop 2023 at the 29th ACM SIGKDD Conference on Knowledge Discovery and Data Mining},
  year      = {2023}
}

@inproceedings{zhao2019unified,
  title     = {A Unified Framework for Marketing Budget Allocation},
  author    = {Zhao, Kui and Hua, Junhao and Yan, Ling and Zhang, Qi and Xu, Huan and Yang, Cheng},
  booktitle = {Proceedings of the 25th ACM SIGKDD International Conference on Knowledge Discovery \& Data Mining (KDD '19)},
  pages     = {3305--3313},
  year      = {2019}
}

@book{kohavi2020trustworthy,
  title     = {Trustworthy Online Controlled Experiments: A Practical Guide to A/B Testing},
  author    = {Kohavi, Ron and Tang, Diane and Xu, Ya},
  year      = {2020},
  publisher = {Cambridge University Press}
}

@article{eckles2017interference,
  title={Design and analysis of experiments in networks: Reducing bias from interference},
  author={Eckles, Dean and Karrer, Brian and Ugander, Johan},
  journal={Journal of Causal Inference},
  volume={5},
  number={1},
  pages={20150021},
  year={2017},
  publisher={De Gruyter}
}

@article{bernal2017its,
  title   = {Interrupted time series regression for the evaluation of public health interventions: a tutorial},
  author  = {Bernal, James Lopez and Cummins, Steven and Gasparrini, Antonio},
  journal = {International Journal of Epidemiology},
  volume  = {46},
  number  = {1},
  pages   = {348--355},
  year    = {2017}
}

@book{angrist2009mhe,
  title     = {Mostly Harmless Econometrics: An Empiricist's Companion},
  author    = {Angrist, Joshua D. and Pischke, J{\"o}rn-Steffen},
  year      = {2009},
  publisher = {Princeton University Press}
}

@article{abadie2021using,
  title   = {Using Synthetic Controls: Feasibility, Data Requirements, and Methodological Aspects},
  author  = {Abadie, Alberto},
  journal = {Journal of Economic Literature},
  volume  = {59},
  number  = {2},
  pages   = {391--425},
  year    = {2021}
}

@article{zhang2019inference,
  title={Inference for Synthetic Control Methods with Multiple Treated Units},
  author={Zhang, Ziyan},
  journal={arXiv preprint arXiv:1912.00568},
  year={2019}
}

@techreport{dube2014pooled_techreport,
  title       = {Pooled Synthetic Control Estimates for Continuous Treatments: An Application to Minimum Wage Case Studies},
  author      = {Dub{\'e}, Arindrajit and Zipperer, Ben},
  institution = {IZA},
  year        = {2014}
}

@book{pearl2009causality,
  title     = {Causality: Models, Reasoning, and Inference},
  author    = {Pearl, Judea},
  edition   = {2},
  year      = {2009},
  publisher = {Cambridge University Press}
}

@article{shimizu2006linear,
  title   = {A Linear Non-Gaussian Acyclic Model for Causal Discovery},
  author  = {Shimizu, Shohei and Hoyer, Patrik O. and Hyv{\"a}rinen, Aapo and Kerminen, Antti},
  journal = {Journal of Machine Learning Research},
  volume  = {7},
  pages   = {2003--2030},
  year    = {2006}
}

@inproceedings{hoyer2009nonlinear,
  title     = {Nonlinear Causal Discovery with Additive Noise Models},
  author    = {Hoyer, Patrik O. and Janzing, Dominik and Mooij, Joris M. and Peters, Jonas and Sch{\"o}lkopf, Bernhard},
  booktitle = {Advances in Neural Information Processing Systems},
  volume    = {21},
  year      = {2009}
}

@book{peters2017elements,
  title     = {Elements of Causal Inference: Foundations and Learning Algorithms},
  author    = {Peters, Jonas and Janzing, Dominik and Sch{\"o}lkopf, Bernhard},
  year      = {2017},
  publisher = {MIT Press}
}

@inproceedings{immer2023identifiability,
  title     = {On the Identifiability and Estimation of Causal Location-Scale Noise Models},
  author    = {Immer, Alexander and Schultheiss, Christoph and Vogt, Julia E. and Sch{\"o}lkopf, Bernhard and B{\"u}hlmann, Peter and Marx, Alexander},
  booktitle = {International Conference on Machine Learning},
  pages     = {14316--14332},
  year      = {2023}
}

@inproceedings{yin2024effective,
  title     = {Effective Causal Discovery under Identifiable Heteroscedastic Noise Model},
  author    = {Yin, Naiyu and Gao, Tian and Yu, Yue and Ji, Qiang},
  booktitle = {Proceedings of the AAAI Conference on Artificial Intelligence},
  volume    = {38},
  number    = {15},
  pages     = {16486--16494},
  year      = {2024}
}

@inproceedings{abbasi2011regret,
  title     = {Regret Bounds for the Adaptive Control of Linear Quadratic Systems},
  author    = {Abbasi-Yadkori, Yasin and Szepesv{\'a}ri, Csaba and Bartlett, Peter},
  booktitle = {COLT},
  year      = {2011}
}

@inproceedings{dean2018regret,
  title     = {Regret Bounds for Robust Adaptive Control of Linear Quadratic Systems},
  author    = {Dean, Sarah and Mania, Horia and Matni, Nikolai and Recht, Benjamin and Tu, Stephen},
  booktitle = {Advances in Neural Information Processing Systems},
  year      = {2018}
}

@inproceedings{srinivas2010gaussian,
  title     = {Gaussian Process Optimization in the Bandit Setting: No Regret and Experimental Design},
  author    = {Srinivas, Niranjan and Krause, Andreas and Kakade, Sham and Seeger, Matthias},
  booktitle = {Proceedings of the 27th International Conference on Machine Learning},
  year      = {2010}
}

@article{auer2002using,
  title={Using confidence bounds for exploitation-exploration trade-offs},
  author={Auer, Peter},
  journal={Journal of machine learning research},
  volume={3},
  number={Nov},
  pages={397--422},
  year={2002}
}

@article{jaksch2010near,
  title   = {Near-optimal Regret Bounds for Reinforcement Learning},
  author  = {Jaksch, Thomas and Ortner, Ronald and Auer, Peter},
  journal = {Journal of Machine Learning Research},
  volume  = {11},
  pages   = {1563--1600},
  year    = {2010}
}

@article{bottou2013counterfactual,
  title   = {Counterfactual reasoning and learning systems: The example of computational advertising},
  author  = {Bottou, L{\'e}on and Peters, Jonas and Qui{\~n}onero-Candela, Joaquin and Charles, Denis X. and Chickering, D. Max and Portugaly, Elon and Ray, Dipankar and Simard, Patrice and Snelson, Ed},
  journal = {The Journal of Machine Learning Research},
  volume  = {14},
  number  = {1},
  pages   = {3207--3260},
  year    = {2013}
}

@book{savage1954foundations,
  title     = {The Foundations of Statistics},
  author    = {Savage, Leonard J.},
  year      = {1954},
  publisher = {Wiley}
}

@article{loomes1982regret,
  title   = {Regret theory: An alternative theory of rational choice under uncertainty},
  author  = {Loomes, Graham and Sugden, Robert},
  journal = {The Economic Journal},
  volume  = {92},
  number  = {368},
  pages   = {805--824},
  year    = {1982}
}
